\documentclass[conference]{IEEEtran}
\IEEEoverridecommandlockouts
\usepackage{cite}
\usepackage{amsmath,amssymb,amsfonts}
\usepackage{graphicx}
\usepackage{textcomp}
\usepackage{subfigure}
\usepackage{color}
\usepackage{algorithm}
\usepackage{algorithmic}
\usepackage{bm}
\usepackage{flushend}

\renewcommand{\algorithmicensure}{\textbf{Output:}}

\graphicspath{{./eps/}}

\def\BibTeX{{\rm B\kern-.05em{\sc i\kern-.025em b}\kern-.08em
    T\kern-.1667em\lower.7ex\hbox{E}\kern-.125emX}}
\begin{document}
\title{Network-Density-Controlled Decentralized Parallel Stochastic Gradient Descent in Wireless Systems}
\author{
Koya Sato\IEEEauthorrefmark{1}, Yasuyuki Satoh\IEEEauthorrefmark{2}, and Daisuke Sugimura\IEEEauthorrefmark{3}\\
\IEEEauthorrefmark{1}\IEEEauthorrefmark{2}Tokyo University of Science, 125-8585, Tokyo, Japan\\
\IEEEauthorrefmark{3} Tsuda University, 187-8577, Tokyo, Japan\\
\IEEEauthorrefmark{1}k\_sato@ieee.org, \IEEEauthorrefmark{2}ysatoh@ee.kagu.tus.ac.jp, \IEEEauthorrefmark{3}sugimura@tsuda.ac.jp
}

\maketitle

\begin{abstract}
    This paper proposes a communication strategy for decentralized learning on wireless systems. Our discussion is based on the decentralized parallel stochastic gradient descent (D-PSGD), which is one of the state-of-the-art algorithms for decentralized learning. The main contribution of this paper is to raise a novel open question for decentralized learning on wireless systems: there is a possibility that the density of a network topology significantly influences the runtime performance of D-PSGD.
    In general, it is difficult to guarantee delay-free communications without any communication deterioration in real wireless network systems because of path loss and multi-path fading. These factors significantly degrade the runtime performance of D-PSGD. To alleviate such problems, we first analyze the runtime performance of D-PSGD by considering real wireless systems.
    This analysis yields the key insights that dense network topology (1) does not significantly gain the training accuracy of D-PSGD compared to sparse one, and (2) strongly degrades the runtime performance because this setting generally requires to utilize a low-rate transmission. Based on these findings, we propose a novel communication strategy, in which each node estimates optimal transmission rates such that communication time during the D-PSGD optimization is minimized under the constraint of network density, which is characterized by radio propagation property. The proposed strategy enables to improve the runtime performance of D-PSGD in wireless systems. 
    Numerical simulations reveal that the proposed strategy is capable of enhancing the runtime performance of D-PSGD.
\end{abstract}


\section{Introduction}
\label{sect:intro}
Based on the rapid development of deep neural networks (DNNs), many machine learning techniques have been proposed over the past decade. In general, constructing an accurate DNN incurs high computational costs and requires massive numbers of training samples. This problem has motivated many researchers to investigate machine learning techniques exploiting distributed computing resources, such as multiple graphical processing units in one computer, multiple servers in a data center, or smartphones distributed over a city\;\cite{Shi-JIoT2016, M-Dong-IEEENetwork2018}. If one can efficiently utilize distributed computation resources, classifiers (or regressors) can be trained in a shorter time period compared to utilizing one machine with single-thread computation.
\par
Several researchers have proposed algorithms for distributed machine learning \cite{Zinkevich2009, Stich-ICLR2019, pmlr-v54-mcmahan17a, DBLP:journals/corr/abs-1902-00340, Lian-NIPS2017,Wang2018-CooperativeSGD}.
According to these past studies, we can categorize distributed machine learning techniques into (a) centralized \cite{Zinkevich2009, Stich-ICLR2019, pmlr-v54-mcmahan17a}, and (b) decentralized settings \cite{DBLP:journals/corr/abs-1902-00340, Lian-NIPS2017, Wang2018-CooperativeSGD}.
\par
The centralized algorithms assume to prepare a centralized server, and all the nodes can connect to this server. Generally, centralized algorithms construct more accurate classifiers compared to decentralized algorithms because a centralized server allows such algorithms to exploit the conditions of all computation nodes (e.g., number of datasets, computational capabilities, and network status), facilitating the construction of an optimal learning strategy.
However, applications of centralized algorithms are restricted to specific situations, such as federated learning\;\cite{pmlr-v54-mcmahan17a, niknam2019federated,Nishio-ICC2019}, because all nodes must communicate with the centralized server.
In contrast, decentralized algorithms enable these systems to construct a classifier in a distributed manner over the local wireless network, thereby facilitating novel applications of machine learning such as image recognition in cooperative autonomous driving\;\cite{Ye-VehMag2018} and the detection of white space in spectrum sharing systems\;\cite{Bkassiny-COMST2013}, {\it without any clouds and edge computing servers}. Towards exploring further applicabilities of distributed machine learning, this paper studies decentralized learning algorithms on wireless systems.  

\subsection{Problem of Decentralized Learning in Wireless Systems}
There is a crucial problem that must be considered to realize decentralized machine learning on wireless network systems.
Existing algorithms for decentralized machine learning\;\cite{DBLP:journals/corr/abs-1902-00340, Lian-NIPS2017, Wang2018-CooperativeSGD} mainly consist of the following two steps: (1)\,updating local models and (2)\,communicating between nodes. In the procedure for local model updating, each computation node refines the model parameters of the classifier to be trained utilizing its own dataset (specific training samples at each computation node). During the communication procedure, the updated model parameters are shared between neighboring nodes. These procedures are performed iteratively until training loss converges. However, the communication procedure tends to be a bottleneck in terms of runtime performance because the number of model parameters that must be communicated is often enormous (e.g., VGG16\;\cite{DBLP:journals/corr/SimonyanZ14a} requires more than 100 million model parameters).
Furthermore, in wireless systems, the communication time required to guarantee successful communication tends to increase based on path loss and multipath fading\;\cite{Goldsmith}. These factors significantly deteriorate the runtime performance of machine learning.
\par
\begin{figure}[t]
  \centering
  \includegraphics[width=95mm,clip]{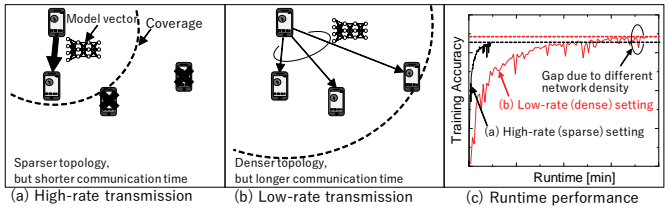}
  \vspace{-13mm}
  \caption{Tradeoffs between transmission rate, network density, and communication time. Past studies\;\cite{Lian-NIPS2017,Wang2018-CooperativeSGD} have shown that the upper bounds on training accuracy for decentralized learning algorithms depends on the density of network topologies. (a) High-rate transmission leads to shorter communication time between nodes, but it can make the network topology sparse, thereby degrading the training accuracy of the classifier\;\cite{Lian-NIPS2017,Wang2018-CooperativeSGD}. (b) Low-rate transmission allows us to facilitate the construction of a dense network topology, resulting in a more accurate classifier, but this strategy requires longer communication times. (c) A numerical example of runtime performance of training accuracy. It clearly shows the tradeoffs between the training accuracy and the runtime performance of decentralized learning.}
  \vspace{-2mm}
  \label{fig:example-strategy}
\end{figure}

This problem is challenging, but should be addressed utilizing either lower or higher transmission rates. Let us consider the situation where the transmitter can controlls the communication coverage by adjusting the transmission rate under given transmission power and bandwidth (e.g., Wi-Fi with adaptive modulation techniques). In general, high-rate transmission can easily reduce communication time. However, this strategy reduces communication coverage, meaning the network topology becomes sparse.
Some theoretical works\;\cite{Lian-NIPS2017,Wang2018-CooperativeSGD} have argued that the training accuracy of decentralized algorithms deteriorates in a sparse network topology. In contrast, low-rate transmission makes network topologies denser, meaning training accuracy versus the number of iterations can be improved. However, runtime performance deteriorates because total communication time increases. We summarize these relationships in Fig.\;\ref{fig:example-strategy}(a)(b), and the tradeoffs between training accuracy and runtime performance that are raised by the differences in the network topology, in Fig.\;\ref{fig:example-strategy}(c).
\par
Therefore, it is important to develop a communication strategy for decentralized learning in wireless systems that improves runtime performance.

\subsection{Objective of This Paper}
In this paper, we analyze the performance of decentralized learning by considering the influences of network topology on wireless systems and propose a novel communication strategy for improving runtime performance.
We specifically focus on decentralized parallel stochastic gradient descent (D-PSGD)\;\cite{Lian-NIPS2017}, which is one of the state-of-the-art algorithms for decentralized learning, as a reference algorithm for our discussion.
Wang {\it et al.}\,\cite{Wang2018-CooperativeSGD} formulated a relationship between network density and the performance of D-PSGD. They analyzed the performance of D-PSGD from the perspective of computation of the average squared gradient norm of a learning model, which directly affects training accuracy.
Based on this analysis, we first discuss when and how network density affects the runtime performance of D-PSGD. This discussion yields the following two insights: dense network topology (1) does not significantly gain the training accuracy of D-PSGD compared to sparse one, and (2) strongly degrades the runtime performance because this setting generally requires to utilize a low-rate transmission.
These insights suggest that the runtime performance of D-PSGD can be improved by high-rate transmission, which makes the network topology relatively sparse, but shortens the communication time between nodes\;(e.g., Fig.\;\ref{fig:example-strategy}(c)).
Motivated by these insights, we propose a communication strategy that makes each node high-rate transmissions whenever possible.
In this method, each node adapts its transmission rate such that the required time for model sharing is minimized under a constraint on network topology density. By increasing the transmission rate without making the network less dense than necessary, this method improves runtime performance while maintaining training accuracy.
To the best of our knowledge, this work is the first attempt that incorporated characteristics of wireless channels into the D-PSGD algorithm in wireless systems.
\section{System Model}
\label{sect:systemmodel}
\subsection{Overview of D-PSGD}
\label{subsec:systemmodel-d-psgd}
Consider situation in which $n$ nodes are randomly deployed in a two-dimensional area.
The $i$-th node stores independent and identically distributed datasets that follow the probability distribution $D$, and has the $N$-dimensional model parameter vector ${\bm x}_i \in \mathbb{R}^N$ of the classifier (or regressor) that consists of the data size $M$\;[bits]. We assume that each node location has been preliminarily shared with all nodes via periodic short-length communication (e.g., beaconing). 
Additionally, we also assume that all nodes can be roughly (ms order) synchronized once the aforementioned periodic short length communication or global positioning system is deployed.
\par
The objective of distributed learning in a decentralized setting is to optimize the model vector. According to \cite{Lian-NIPS2017}, this objective can be modeled as
\begin{equation}
 \min_{{\bm x}_1, {\bm x}_2, \cdots, {\bm x}_n} \frac{1}{n}\sum^{n}_{i=1} \mathbb{E}_{\xi\thicksim D} \left[F_i\left({\bm x};\xi\right)\right],
  \label{eq:obj-d-psgd}
\end{equation}
where ${\bm x} \triangleq \frac{1}{n}\sum_{i=1}^n {\bm x}_i$, $\xi$ denotes the data sample and $F_i$ represents the loss function for the $i$-th node. After the optimization, each node can utilize ${\bm x}_i$ as its classifier. Note that ${\bm x}$ is not directly calculated during the optimization.
\par
Under the conditions described above, decentralized learning can be performed utilizing a D-PSGD optimizer. D-PSGD iteratively performs the following procedure until the value of the loss function is minimized: (1) updating the model parameter ${\bm x}_i$ at each node based on its dataset with the learning rate $\eta$, (2) sharing updated model parameters with connected neighboring nodes, and (3) averaging received and own model parameters. The pseudo-code of this algorithm is summarized in Algorithm\;\ref{alg:d-psgd}. In Algorithm\;\ref{alg:d-psgd}, we denote the set of model vectors at the $k$-th iteration as ${\bm X}_k = \left({\bm x}_{k, 1},{\bm x}_{k, 2},\cdots, {\bm x}_{k, n}\right)$.

\begin{algorithm}[t]
  \caption{D-PSGD on the $i$-th node\;\cite{Lian-NIPS2017}}
  \label{alg:d-psgd}
  \begin{algorithmic}[1]
    \REQUIRE initial point ${\bm x}_{0,i} = {\bm x}_0$, learning rate $\eta$, and number of iterations $K$.
    \FOR{$k=0,1,2,\cdots,K-1$}
    \STATE{Randomly sample $\xi_{k,i}$ from local data of the $i$-th node.}
    \STATE{Broadcast and receive model parameters to/from neighboring nodes.}
    \STATE{Calculate intermidiate model ${\bm x}_{k+\frac{1}{2},i}$ by averaging the received and own models}
    \STATE{Update the local model parameters ${\bm x}_{k+1,i}\leftarrow {\bm x}_{k+\frac{1}{2},i}-\eta \nabla F_i({\bm x}_{k,i};\xi_{k,i})$.}
    \ENDFOR
  \end{algorithmic}
\end{algorithm}

\subsection{Radio Propagation Model and Protocol}
\label{subsec:systemmodel-propagation}
In wireless systems, the communication coverage is strongly affected by the relationships between the radio propagation characteristics, bandwidth, transmission rate, etc. In order to discuss the influence of these relationships on the performance of D-PSGD, we consider a typical wireless channel.
\par
Because the communication coverage is mainly determined by the path loss, we model the received signal power at a distance $d$\;[m] as $P(d) = P_\mathrm{Tx} - 10\epsilon\log_{10}d\;\;\mathrm{[dBm]}$,
where $P_\mathrm{Tx}$ is the transmission power in dBm and $\epsilon$ is the path loss index. We assume that all nodes transmit with the same $P_\mathrm{Tx}$ and the bandwidth $B$. Under these conditions, the channel capacity at $d$ can be expressed as
\begin{equation}
  C(d) = B \log_{2}\left(1+\frac{\gamma (d)}{B}\right)\;\;\mathrm{[bps]},
  \label{eq:channelcapacity}
\end{equation}
where $\gamma (d)=10^{\frac{P(d)-N_0}{10}}$ is the signal-to-noise ratio and $N_0$ is the noise floor in dBm.
Additionally, we define $n\times n$ channel-capacity matrix ${\bm C}$ whose element $C_{ij}$ represents the channel capacity between the $i$-th and the $j$-th nodes. 
\par
This paper assumes situations where each node can controll its communication coverage by adjusting the transmission rate. In such situations, we consider that each node broadcasts its own updated model at a transmission rate $R_i$ [bps] (Step 3 in Algorithm\;\ref{alg:d-psgd}).
If $C(d)\geq R_i$, the receiver can accurately receive the model parameters from neighboring nodes. We assume that $N_0$ and $\epsilon$ are constant over the area and that they can be given as prior knowledge to all nodes.
Additionally, to avoid communication collisions between the nodes in Step 3 of Algorithm\;\ref{alg:d-psgd}, the nodes share the spectrum based on the time division multiplexing; the model parameter ${\bm x}_i$ is broadcasted to nodes in consecutive order from the terminal on the west side of the target area. With these assumptions, the communication time spent in one iteration is given by
\begin{equation}
    t_\mathrm{com} = M\sum_{i=1}^n \frac{1}{R_i}\;\text{[sec/share]}.
    \label{eq:communtime}
\end{equation}
If the transmission power $P_\mathrm{Tx}$ and bandwidth $B$ are constrained, the transmission rate $R_i$ must be reduced to expand communication coverage (i.e., to make the network dense). This fact indicates that there is a tradeoff between network density and communication time when sharing model parameters. Therefore, even if the training accuracy of D-PSGD for a given number of iterations can be improved, runtime performance would deteriorate.
\par
Note that our discussion can be extended to fading channels without loss of generality of our claim. This can be achieved by considering the following condition for successful communications: $R\leq \left(C(d) - \Delta C\right)$, where $\Delta C (\geq 0)$ is a constant scalar that behaves as the margin of uncertainty for fading channels. These conditions enable each node to set a transmission rate $R_i$ to perform accurate communication.

\subsection{Modeling D-PSGD using Averaging Matrix}
Previous studies\;\cite{Lian-NIPS2017,Wang2018-CooperativeSGD} have utilized an averaging matrix ${\bm W} \in \mathbb{R}^{n \times n}$, which is automatically determined based on the network topology, for the analysis of D-PSGD.
This averaging matrix ${\bm W}$ satisfies ${\bm W}{\bm 1} = {\bm 1}$, where ${\bm 1}$ is an $n$-dimensional column vector of ones. Each element $W_{ij}$ can be calculated by
\begin{equation}
    W_{ij} = \frac{A_{ij}}{\sum_{j=1}^n A_{ij}},\;\;
    A_{ij} = 
    \begin{cases}
        1\;\;\text{if } C_{ij} \geq R_i \\
        0\;\;\text{otherwise}
    \end{cases},
    \label{eq:weightmatrix}
\end{equation}
where $A_{ij}$ represents the connectivity between the $i$-th and the $j$-th nodes.

The use of ${\bm W}$ allows us to analyze the influence of network topology on D-PSGD. The model updating rule at the $k+1$\,th iteration (i.e., Step\;5 in Algorithm\;\ref{alg:d-psgd}) can be re-defined as
\begin{equation}
  \begin{pmatrix}
   {\bm x}_{k+1,1} \\
   {\bm x}_{k+1,2} \\
   \vdots \\
   {\bm x}_{k+1,n}
  \end{pmatrix}
  \leftarrow
  {\bm W}
  \begin{pmatrix}
   {\bm x}_{k,1} \\
   {\bm x}_{k,2} \\
   \vdots \\
   {\bm x}_{k,n}
  \end{pmatrix}
  -\eta
  \begin{pmatrix}
   \nabla F_1({\bm x}_{k,1};\xi_{k,1})\\
   \nabla F_2({\bm x}_{k,2};\xi_{k,2})\\
   \vdots \\
   \nabla F_3({\bm x}_{k,n};\xi_{k,n})
  \end{pmatrix}.
  \label{eq:UpdateOfD-PSGD}
\end{equation}
In this paper, we also utilize Eq.\,\eqref{eq:UpdateOfD-PSGD} for analyzing the influence of network topology on the runtime performance of D-PSGD.
\section{Network-Density-Controlled D-PSGD}
\label{sect:proposed}
\subsection{Effects of Network Density}
\label{subsec:effect-of-density}
\begin{figure*}[t]
    \centering
    \subfigure[$K=1$, $n=6$.]{
    \includegraphics[width=42mm,clip]{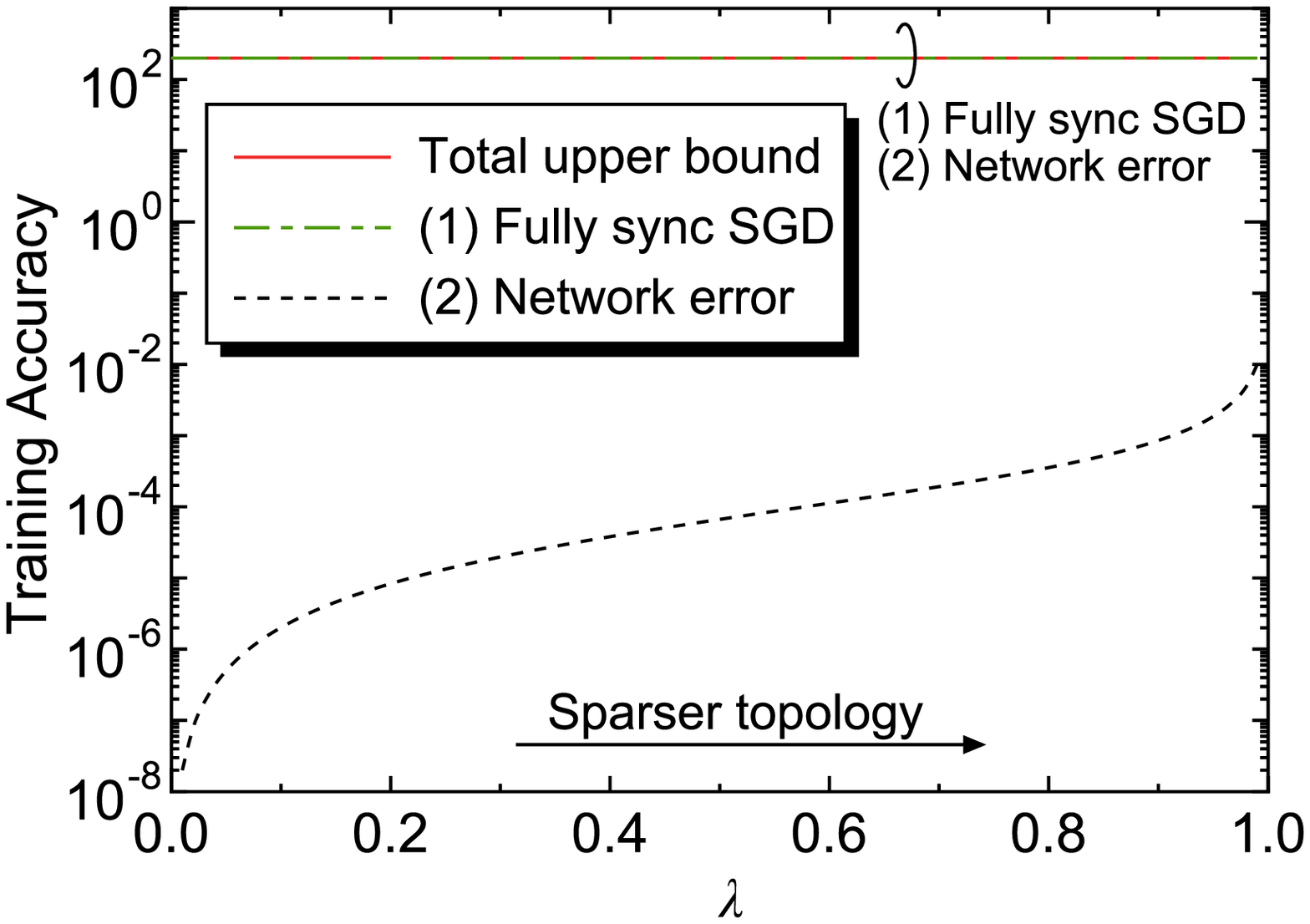}
    }
    \subfigure[$K=100$, $n=6$.]{
    \includegraphics[width=42mm,clip]{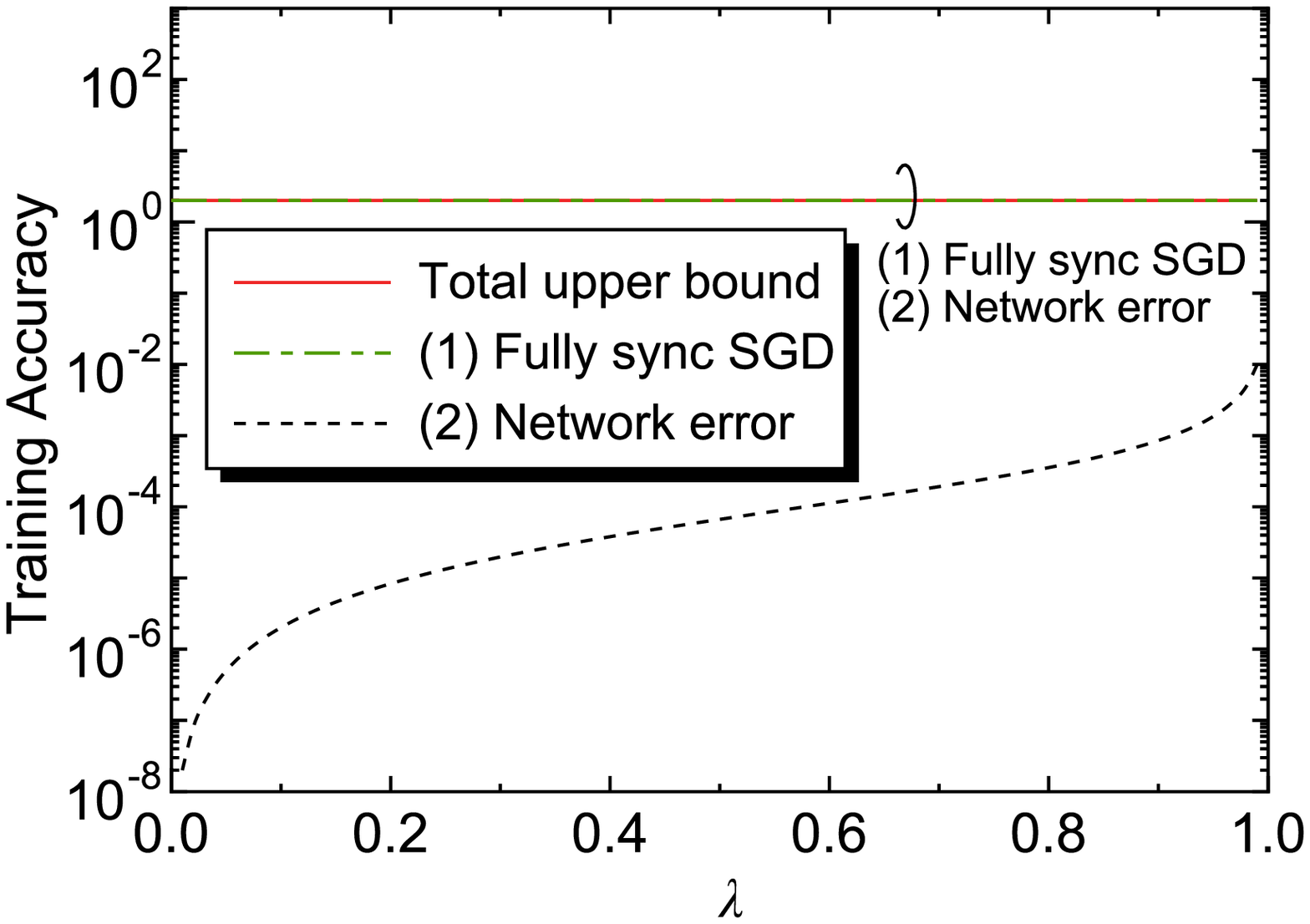}
    }
    \subfigure[$K\rightarrow \infty$, $n=6$.]{
    \includegraphics[width=42mm,clip]{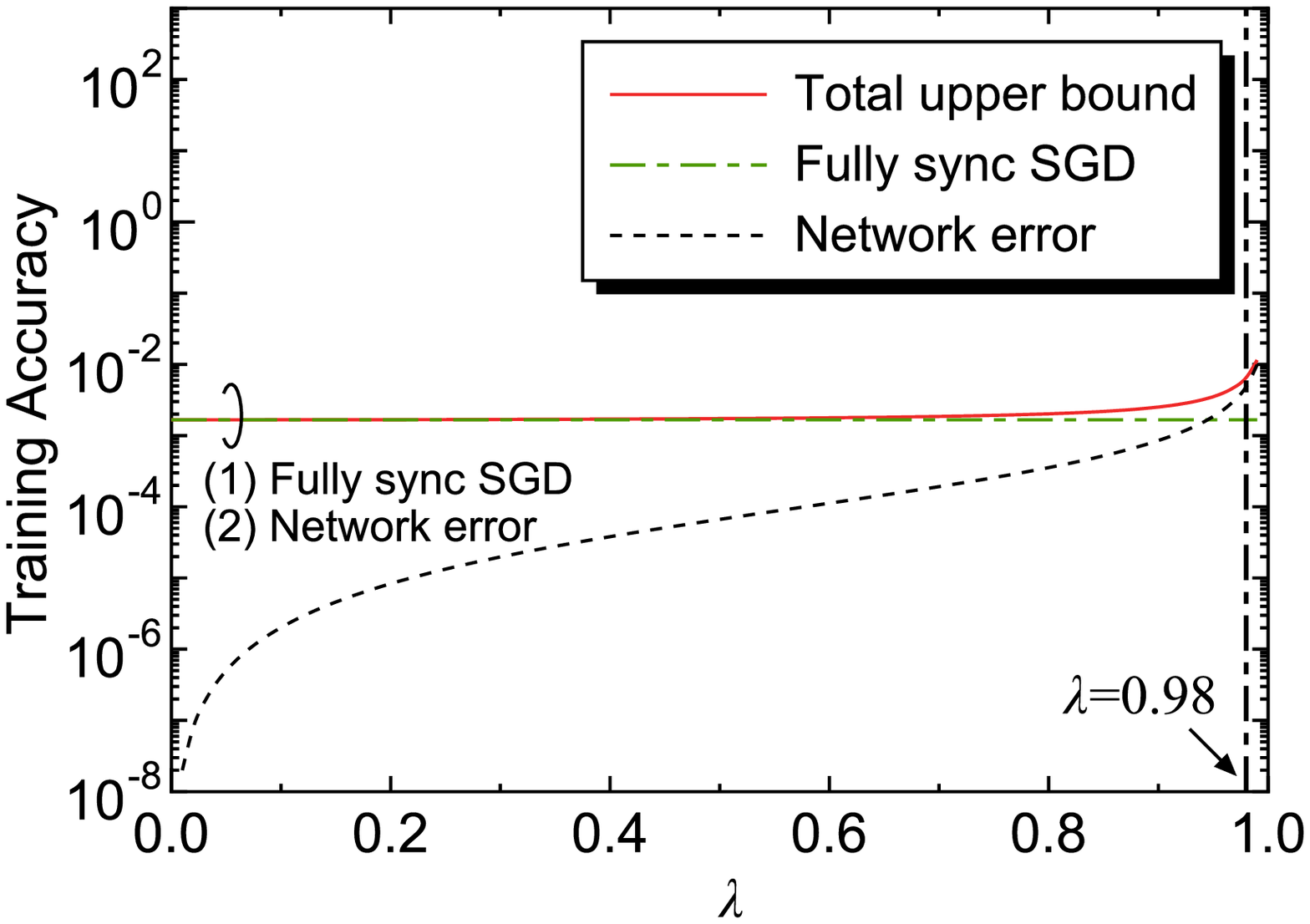}
    }
    \subfigure[Effect of $n$ where $K\rightarrow \infty$.]{
    \includegraphics[width=42mm,clip]{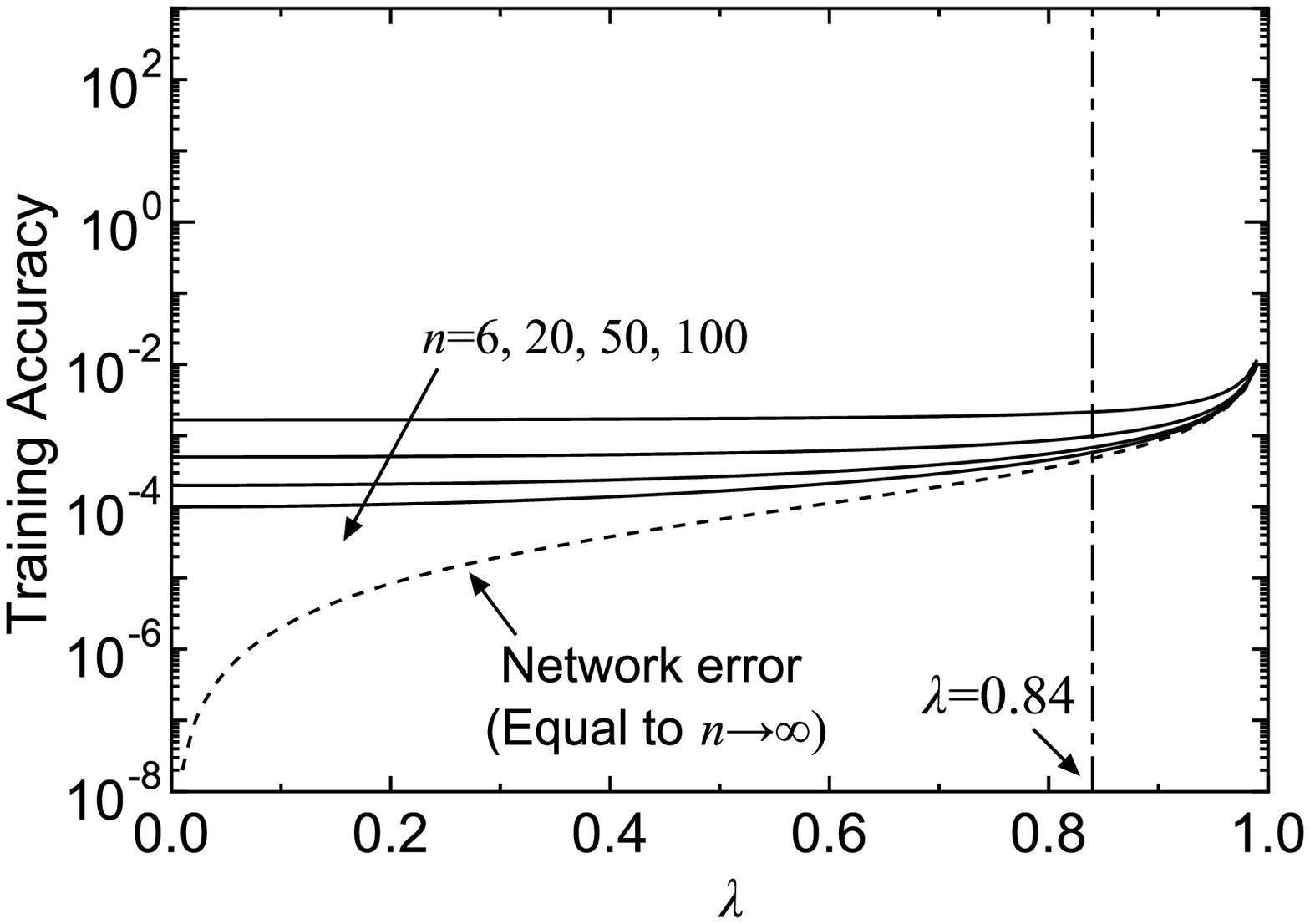}
    }
    \caption{Effects of $\lambda$ on D-PSGD (the Lipschitz constant of the objective function $L=1$, the variance bound of mini-batch SGD $\sigma^2=1$, the learning rate $\eta=0.01$, $F_1 = 1$, and $F_\mathrm{inf}=0$). 
    For various values of $K$ and $n$, if $\lambda$ is below a certain threshold (e.g., $\lambda\leq 0.98$ in (c) and $\lambda\leq 0.84$ in (d) where $n=20$), reducing $\lambda$ does not improve the upper bound significantly, at least on the order level. This numerical example implies that we can boost runtime performance by making the network topology more sparse (i.e., making the transmission rate higher) {\it without significant degradation of training accuracy}.}
    \label{fig:effect-lambda}
\end{figure*}

We will briefly discuss how the density of a network topology influences the training accuracy of D-PSGD.
Wang {\it et al.}\,\cite{Wang2018-CooperativeSGD} analyzed the performance of D-PSGD from the perspective of convergence analysis of the expected value of the squared gradient norm $\mathbb{E}\left[\frac{1}{K}\sum_{k=1}^K || \nabla F({\bm X}_k)||^2\right]$, where $K$ is the number of iterations of optimization for D-PSGD. Because this expected value is directly related to training accuracy, we present this value as ``training accuracy'' throughout this paper.
\par
According to \cite{Wang2018-CooperativeSGD}, the training accuracy of D-PSGD decreases as the parameter $\lambda=\max\left\{|\lambda_2({\bm W})|, |\lambda_{n}({\bm W})|\right\}$ ($\lambda_2({\bm W})$ and $\lambda_n({\bm W})$ are the 2nd and $n$-th largest eigenvalue of ${\bm W}$, respectively) increases. 
The parameter $\lambda$ approaches zero as the number of non-zero elements in ${\bm W}$ increases. This behavior of $\lambda$ suggests that the value of $\lambda$ represents the sparseness of a network topology because a denser network topology causes the number of non-zero elements in ${\bm W}$ to increase.
\par
To derive theoretical proof of D-PSGD performance evaluations, the authors of \cite{Wang2018-CooperativeSGD} introduced the following assumptions:
\begin{itemize}
    \item (Smoothness): $||\nabla F({\bm x}) - \nabla F({\bm y})||\leq L||{\bm x} - {\bm y}||$ ($L$ is the Lipschitz constant of the loss function $F$).
    \item (Lower bounded): $F({\bm x}) \geq F_\mathrm{inf}$.
    \item (Unbiased gradients): $\mathbb{E}_{\xi|{\bm x}}\left[g({\bm x})\right] = \nabla F({\bm x})$ ($g({\bm x})$ is the gradient of ${\bm x}$)
    \item (Bounded variance) $\mathbb{E}_{\xi|{\bm x}}\left[||g({\bm x}) - \nabla F({\bm x})||^2\right] \leq \beta ||\nabla F({\bm x})||^2 + \sigma^2$ ($\beta$ and $\sigma^2$ are non-negative constants that are inversely proportional to the mini-batch size).
    \item (Averaging matrix): $\max\left\{|\lambda_2({\bm W})|, |\lambda_{n}({\bm W})|\right\} < \lambda_1 ({\bm W}) = 1$.
    \item (Learning rate): learning rate $\eta$ should satisfies
    \begin{equation}
   \eta L + 5\eta^2 L^2 \left(\frac{1}{1-\lambda}\right)^2 \leq 1.
   \label{eq:condition-of-analysis}
\end{equation}
\end{itemize}

Under these assumptions, when all local models are initialized with the same vector ${\bm x}_0$, the average squared gradient norm at the $K$-th iteration is bounded by:
\begin{align}
    \mathbb{E}\left[\frac{1}{K}\sum_{k=1}^K || \nabla F({\bm X}_k)||^2\right] \leq \underbrace{\frac{2\left[F({\bm X}_1) - F_\mathrm{inf}\right]}{\eta K}+ \frac{\eta L \sigma^2}{n}}_{\text{(1)\;fully-synchronized SGD}} + \nonumber\\
    \underbrace{\eta^2 L^2 \sigma^2 \left(\frac{1+\lambda^2}{1-\lambda^2} - 1\right)}_{\text{(2)\;network error}}.
    \label{eq:bound}
\end{align}
This equation indicates that the upper bound of the average squared gradient norm can be expressed based on the following two factors.
The first ((1) in Eq.\;\eqref{eq:bound}) is a component obtained from fully-synchronized SGD (i.e., ${\bm W}=\left({\bm 1}{\bm 1}^\top\right)/\left({\bm 1}^\top{\bm 1}\right)$).
The second ((2) in Eq.\;\eqref{eq:bound}) is a component generated by network errors, which are influenced by the density of network topology. The condition in Eq.\;\eqref{eq:bound} implies that training accuracy is strongly affected by $\lambda$, when $K$ and $n$ are large. Therefore, we evaluated effects of these parameters on the training accuracy.
\par
Figs.\,\ref{fig:effect-lambda}(a)-(c) plot three numerical examples of Eq.\,\eqref{eq:bound} where $K=1, 100$, and $K\rightarrow \infty$, respectively.
To highlight the influence of the network topology on the training accuracy of D-PSGD, we plot three curves: the total upper bound (value of the right side of Eq.\,\eqref{eq:bound}), effects of fully-synchronized SGD (value of the term (1) on the right side of Eq.\,\eqref{eq:bound}), and the effect of network errors (value of the term (2) on the right side of Eq.\,\eqref{eq:bound}). 
These examples show that as the number of iterations $K$ increases, the impact of network density on the training accuracy of D-PSGD increases, i.e., the effects of network error turns out being dominant with respect to the training accuracy (upper bound). 
However, the effect is small when the value of $\lambda$ is below a certain threshold.
For example, although the effects of $\lambda$ become significant when $K\rightarrow \infty$, the upper bound in this case is on the order of $10^{-2}$ in all regions where $\lambda \leq 0.98$.
The effect of the number of nodes $n$, where $K\rightarrow \infty$, is presented in Fig.\;\ref{fig:effect-lambda}(d). Although the effect of $\lambda$ on the training accuracy increases as $n$ increases, a similar dependence on $\lambda$ threshold can be observed in this case (e.g., $\lambda\leq 0.84$ in where $n=20$). 
These numerical examples suggest that runtime performance can be improved by making a network topology more sparse (i.e., by increasing the transmission rates of nodes) {\it without a significant degradation in training accuracy}. 

\subsection{Proposed Communication Strategy}
\label{subsec:proposedmethod}
As shown earlier, setting a higher transmission rate under the constraint of the network density will improve the runtime performance. Considering the relationships between transmission rate, network density, communication time, and the training accuracy of D-PSGD, we propose a novel communication strategy. In this strategy, each node selects a suitable transmission rate $R_i$ prior to initiating D-PSGD. Once $R_i$ is determined, each node broadcasts its model vector ${\bm x}_i$ based on the transmission rate $R_i$. This transmission rate is selected, such that communication time $t_\mathrm{com}$ is minimized under constraints with respect to $\lambda$. This strategy can be modeled as
\begin{align}
    &\min_{{\bm R}}\;t_\mathrm{com}\;\text{[sec/share]} \nonumber \\
    &\mathrm{s.t.}\;\;\lambda \leq \lambda_\mathrm{target},
    \label{eq:objectivefunction}
\end{align}
where ${\bm R}=\{R_1, R_2, \cdots, R_n\}$ denotes the set of transmission rates and $\lambda_\mathrm{target}$ represents the predetermined maximum value of $\lambda$ (that satisfies Eq.\;\eqref{eq:condition-of-analysis}). This strategy enables one to increase each transmission rate $R_i$, resulting in a sparse network topology. Because the constraint of $\lambda_\mathrm{target}$ prevents significant degradation of training accuracy, runtime performance can be improved while maintaining training accuracy.
\subsection{Solver for Eq.\;\eqref{eq:objectivefunction}}
Eq.\;\eqref{eq:objectivefunction} should be solved at each node in a decentralized manner. There are some methods for optimizing ${\bm R}$ based on given conditions, such as prior knowledge (i.e., with or without location information) and channel characteristics. This paper considers that both pre-shared information at node locations and path loss characteristics, i.e., the received signal power $P(d)$, the bandwidth $B$ and the noise floor $N_0$ can be obtained beforehand.
With this knowledge, each node can construct the channel-capacity matrix ${\bm C}$ independently. This matrix enables the $i$-th node to estimate the required transmission rate $R_i$ that guarantees successful communications with the $j$-th node. Thus, Eq.\;\eqref{eq:objectivefunction} can be expressed as a $n^n$ combination problem. In this paper, each node solves this problem utilizing a brute force search. We summarize these procedures in Algorithm\;\ref{alg:optimization}. Even if each node solves this problem in a decentralized manner, all nodes arrive at the same result.
\par
After ${\bm R}$ is determined, each node initiates D-PSGD with the optimized transmission rate.

\begin{algorithm}[t]
  \caption{Estimation of Optimal Transmission Rate ${\bm R}$ (Solver for Eq.\;\eqref{eq:objectivefunction})}
  \label{alg:optimization}
  \begin{algorithmic}[1]
    \REQUIRE Transmission power $P_\mathrm{Tx}$, noise floor $N_0$, bandwidth $B$, path loss index $\epsilon$ node locations, and $\lambda_\mathrm{target}$.
    \STATE{Calculate the channel-capacity matrix ${\bm C}$ using Eq.\;\eqref{eq:channelcapacity}.}
    \FOR{all candidates of ${\bm R}$}
    \STATE{Construct a candidate of ${\bm R}$ by selecting one $C_{ij}$ from each row.}
    \STATE{Construct averaging matrix ${\bm W}$ using Eq.\;\eqref{eq:weightmatrix}.}
    \STATE{Calculate $\lambda=\max\left\{|\lambda_2({\bm W})|, |\lambda_{n}({\bm W})|\right\}$.}
    \STATE{Search for ${\bm R}$ that minimizes the communication time $t_\mathrm{com}$ under the constraint $\lambda \leq \lambda_\mathrm{target}$.}
    \ENDFOR
    \RETURN{optimized ${\bm R}$}
  \end{algorithmic}
\end{algorithm}
\section{Performance Evaluation}
\label{sect:performance}
  We simulated the proposed strategy on a computer employing a multi-core CPU.
  This computer employs AMD Ryzen Threadripper 2970WX, which consists of 24-physical cores\footnote{Simultaneous multi-threading (SMT) was disabled.}, and works with Ubuntu 18.04 LTS. The simulation program was implemented with PyTorch 1.0.1 on Python 3.7.3.
  \par
  We conducted simulations of a case where six nodes are placed in a 200\;m$\times$200\;m area as shown in Fig.\;\ref{fig:nodeplacement}. We focus on the training accuracy at Node 1.

\subsection{Experimental Setup}
We evaluated the proposed strategy on an image classification task utilizing the Fashion-MNIST dataset\;\cite{xiao2017fashionmnist}, which has been widely used as a benchmark for image classification performance in the machine learning community. This dataset includes 60\,000 images for training that have already been categorized into ten different categories. This dataset also includes 10\,000 images for test data. Each sample in this dataset is a single-channel, 8-bit image with a resolution of $28\times 28$. In this experiment, we utilized a convolutional neural network (CNN) as an architecture to perform image classification. The details of the CNN we utilized are as follows: two convolutional layers (with 10 and 20 channels,  respectively, each of which was activated by a rectified linear unit (ReLU) function), two $2\times 2$ max-pooling layers, and three fully-connected layers (320 and 50 units, respectively, with ReLU activation and an additional 10 units activated by the softmax function). Additionally, dropout was applied in the second convolutional layer and the first fully-connected layer with a dropout ratio of 0.5. Therefore, the total number of model parameters for the CNN was 21\,840, and its data size was $M$=698\,880‬ bits (32-bit floating point numbers). Each node broadcasted data to neighboring nodes to train the CNN utilizing D-PSGD.
To train a CNN utilizing the D-PSGD optimizer, we shuffled all of the training samples, then equally distributed them to six computation nodes. Therefore, each node was given 10\;000 independently and identically distributed training samples. Additionally, we set the batch size for D-PSGD optimization to 1, meaning the number of iterations per epoch was $K=10^4$, because each node was given 10\,000 training samples.
\par
We exected $n$ processes in parallel to train the CNN with D-PSGD on the computer, where we assigned one physical core to each process. The runtime of the calculation portion of D-PSGD was calculated based on the real elapsed time on the computer, and the communication time was calculated by Eq.\;\eqref{eq:communtime}. Note that we fixed a random seed at the start of the simulation to ensure reproducibility.

\subsection{Runtime Performance Results}
In this section, we discuss the experimental results for the proposed strategy in terms of runtime performance. 
\par
We analyzed the performance of the proposed strategy by varying the path loss index $\epsilon$ because communication coverage, which is a key factor influencing runtime performance, is strongly affected by path loss. The path loss index $\epsilon$ is an environment-dependent factor that has been determined empirically. It tends to take on large values in environments with many obstacles, e.g., indoor and urban channels\,\cite{Goldsmith}.
\par
\begin{figure}[t]
    \centering
    \subfigure[Node placement.]{
    \includegraphics[width=38mm,clip]{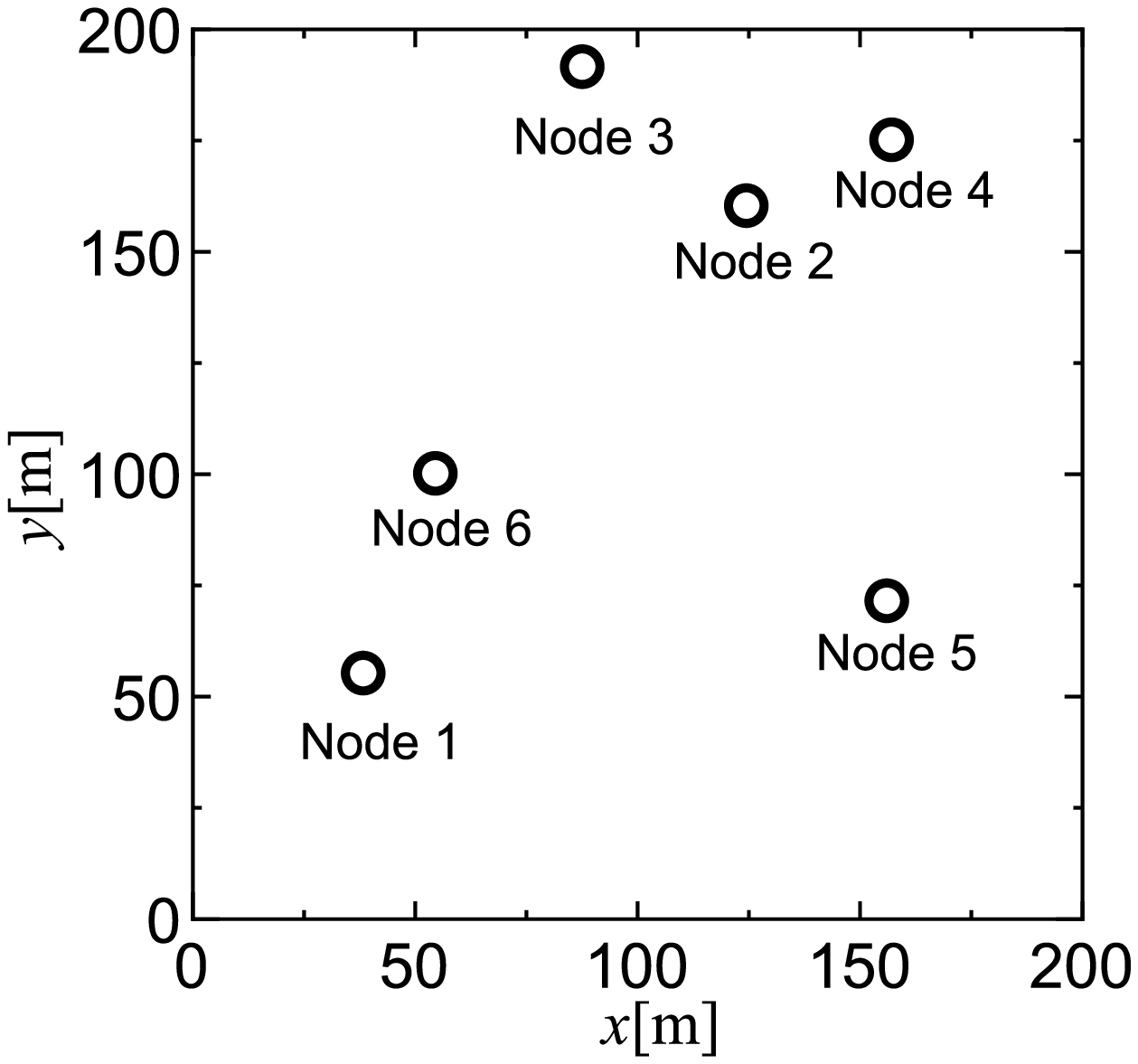}
    \label{fig:nodeplacement}
    }
    \subfigure[Epoch ($\epsilon=3,4,5,6$).]{
    \includegraphics[width=37mm,clip]{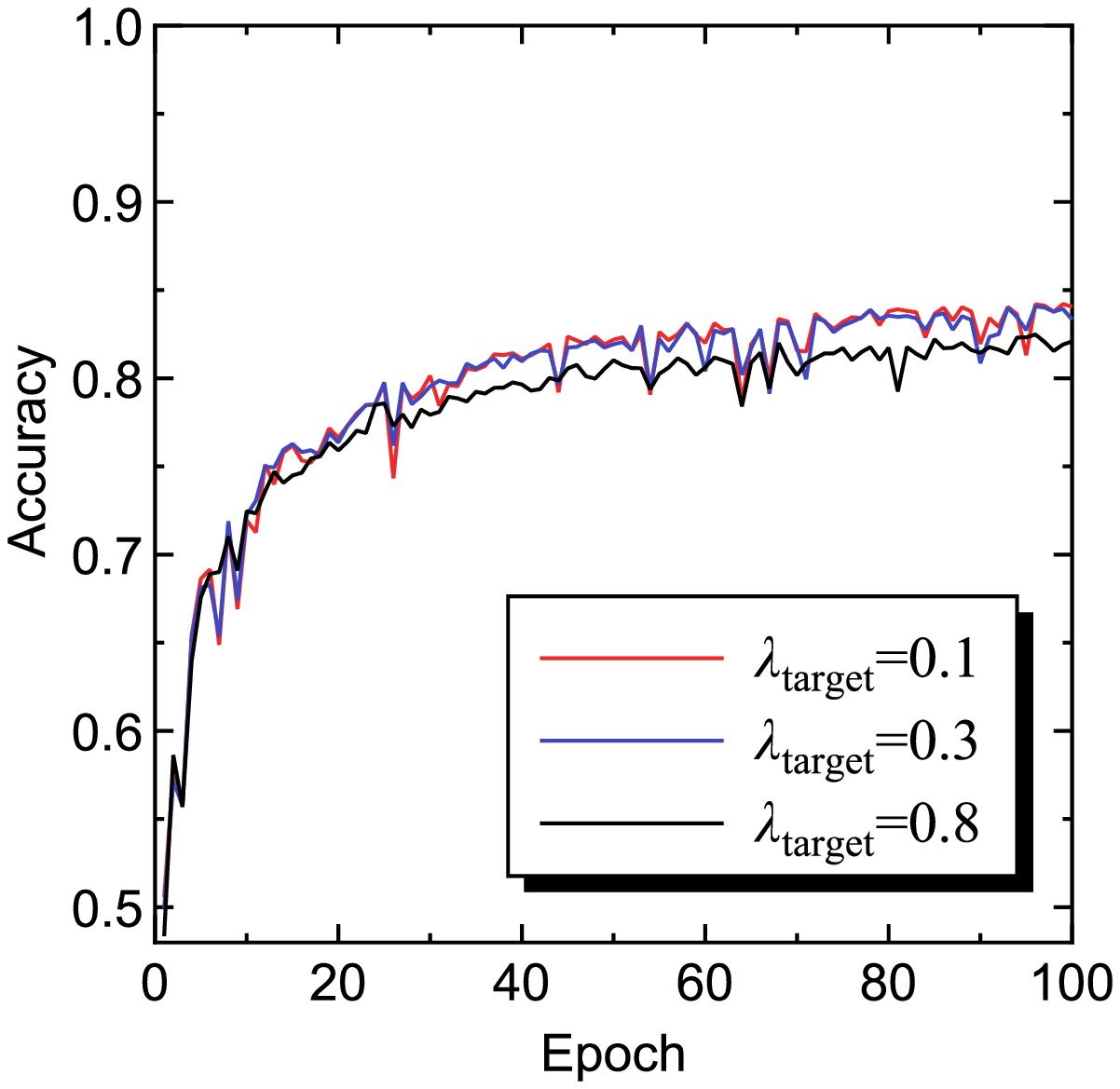}
    \label{fig:epoch}
    }
    \subfigure[Runtime ($\epsilon=3$).]{
    \includegraphics[width=37mm,clip]{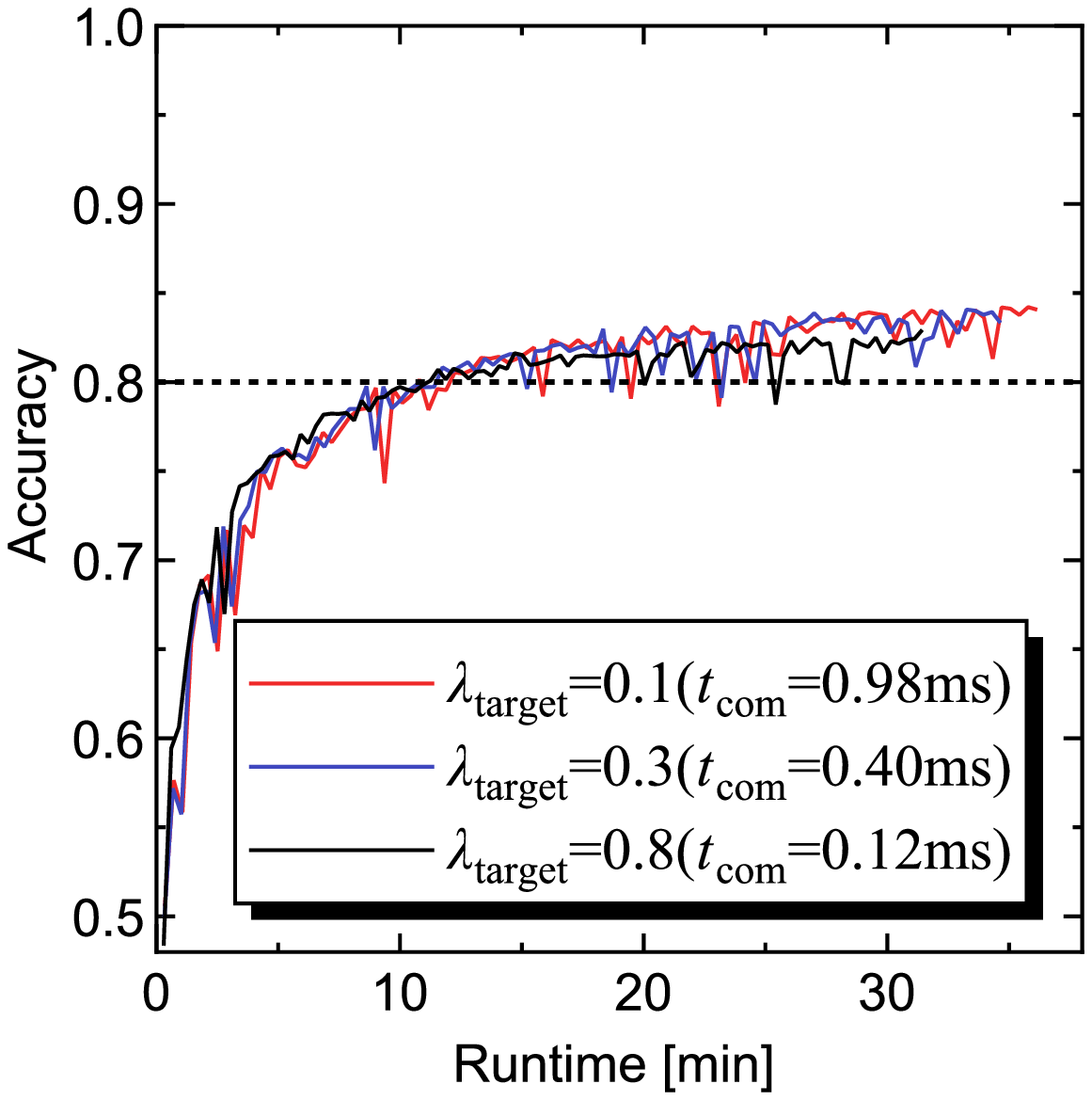}
    \label{fig:eta3-runtime}
    }
    \subfigure[Runtime ($\epsilon=4$).]{
    \includegraphics[width=37mm,clip]{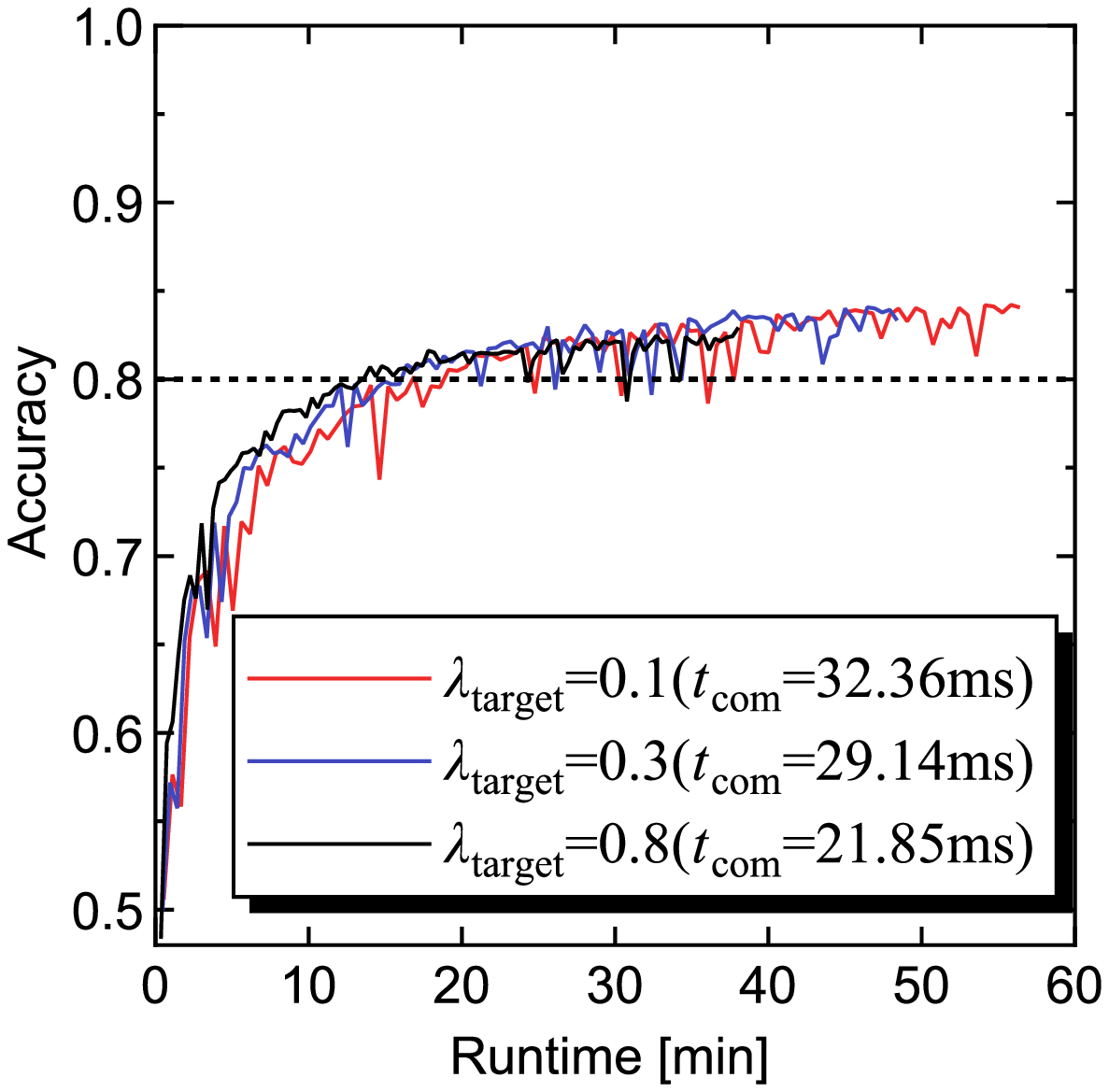}
    \label{fig:eta4-runtime}
    }
    \subfigure[Runtime ($\epsilon=5$).]{
    \includegraphics[width=37mm,clip]{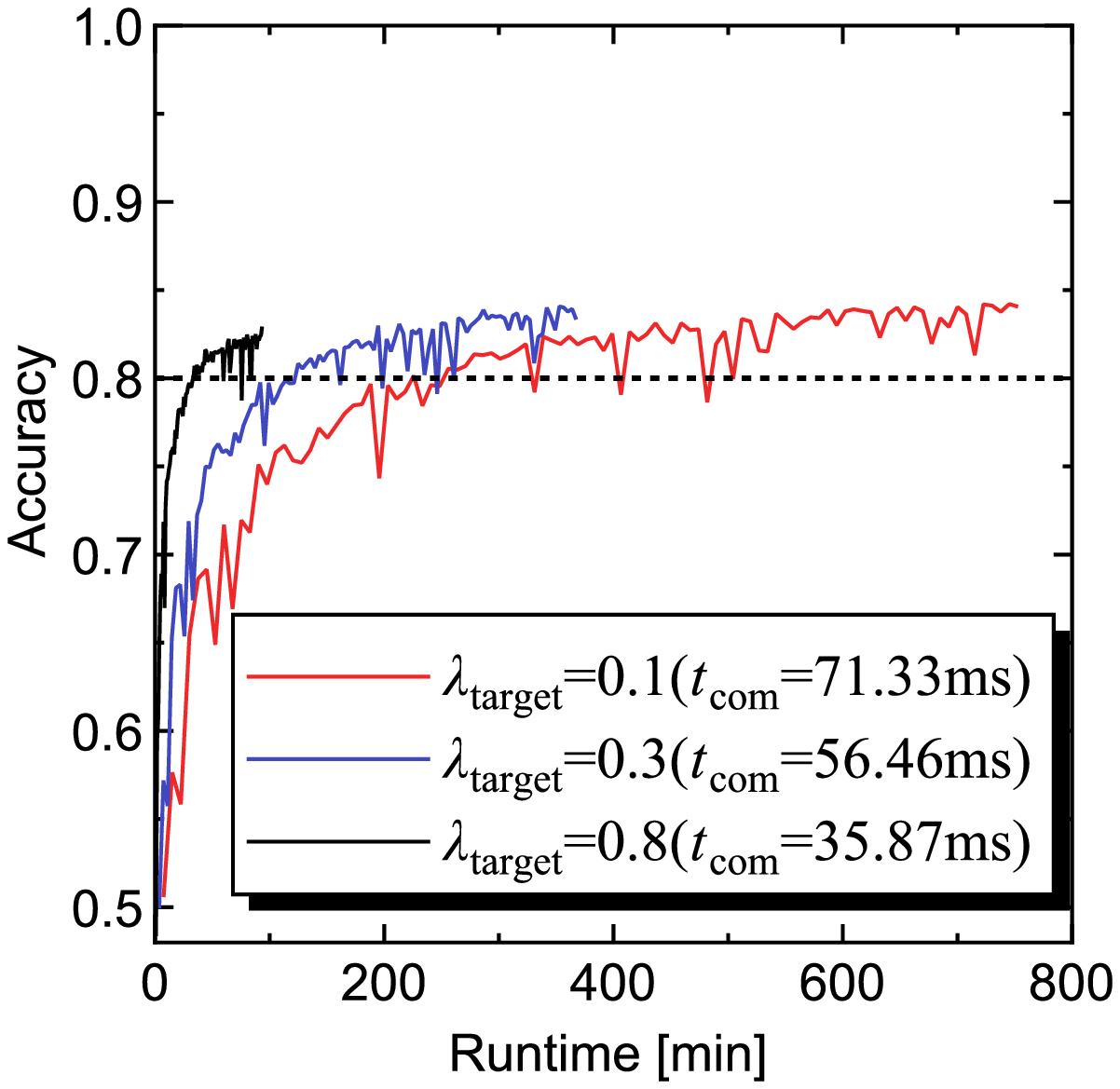}
    \label{fig:eta5-runtime}
    }
    \subfigure[Runtime ($\epsilon=6$).]{
    \includegraphics[width=37mm,clip]{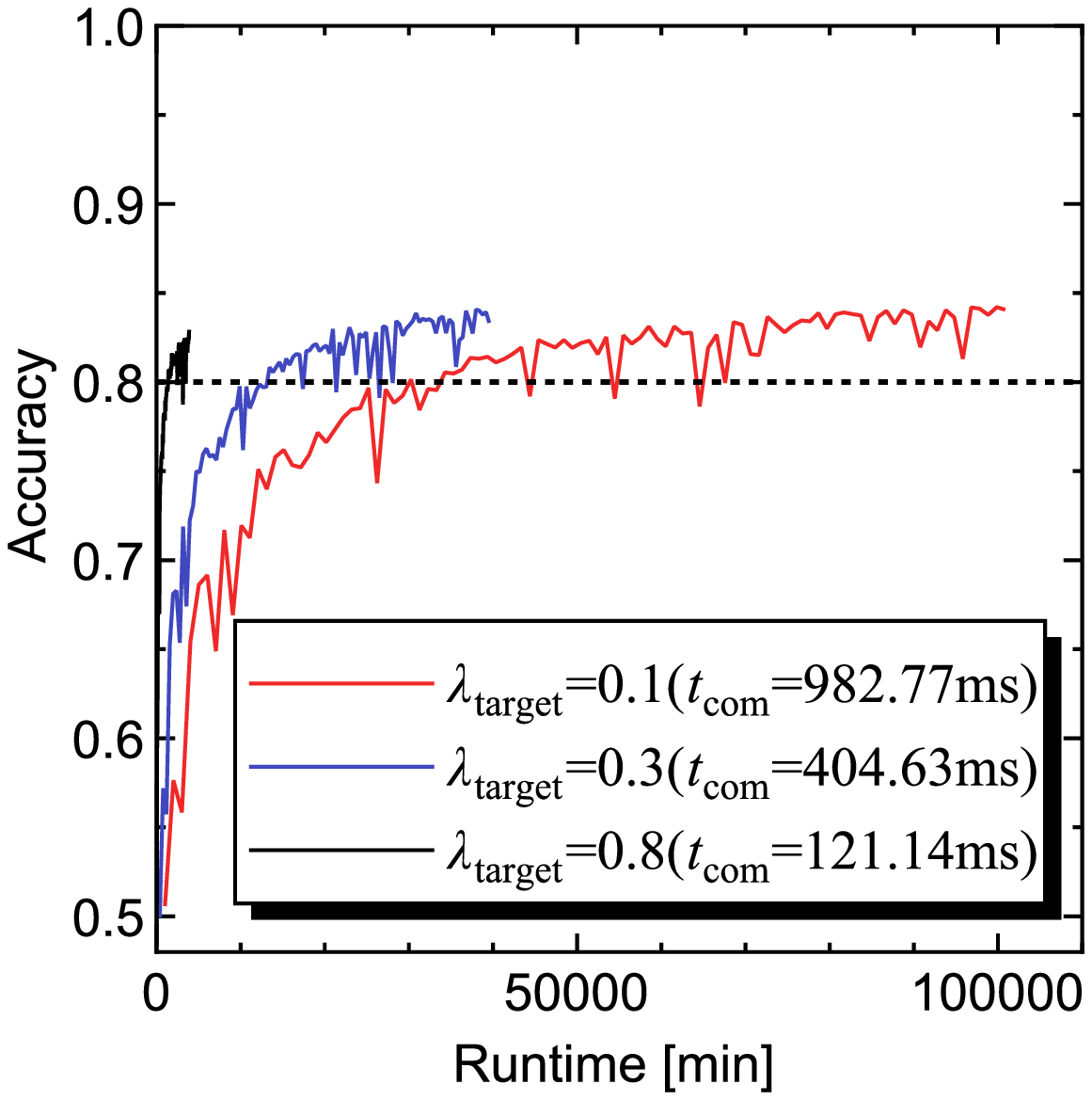}
    \label{fig:eta6-runtime}
    }
    \vspace{-2mm}
    \caption{Training accuracy at Node\;1 (transmission power $P_\mathrm{Tx}=0\;\text{[dBm]}$, bandwidth $B=20\;\text{[MHz]}$, noise floor $N_0=-172.0\;\text{[dBm/Hz]}$, and learning rate $\eta = 0.01$). Although $\lambda_\mathrm{target}$ has almost no effect on epoch performance, a greater value of $\lambda_\mathrm{target}$ clearly improves runtime performance, especially in situations where the path loss index $\epsilon$ is large.}
    \vspace{-5mm}
    \label{fig:result-cnn}
\end{figure}

Fig.\,\ref{fig:result-cnn}(a) present dependences of the training accuracy against the number of epochs for $\epsilon = 3, 4, 5,$ and 6, respectively.
We highlight examples of the obtained training accuracy values at 100 epochs: 0.841 ($\lambda_\mathrm{target} = 0.1$), 0.833 ($\lambda_\mathrm{target} = 0.3$), and 0.821 ($\lambda_\mathrm{target} = 0.8$). These results indicate that the training accuracy decreases slightly as $\lambda_\mathrm{target}$ increases. They agree with the theoretical and numerical evaluations of the performance of D-PSGD in Fig.\;\ref{fig:effect-lambda} and Eq.\;\eqref{eq:bound}. Note that this epoch performance does {\it not} depend on $\epsilon$ because the proposed method always constructs the same network topology for a given $\lambda_\mathrm{target}$ and node placements, regardless of $\epsilon$.
\par
Figs.\,\ref{fig:result-cnn}(c)-(f) present the runtime performances for $\epsilon=3,4,5,$ and $6$, respectively.
When $\epsilon$ is large, a greater value of $\lambda_\mathrm{target}$ (i.e., the higher transmission rate and sparse network topology) significantly improves runtime performance, although this strategy degrades the training accuracy versus epoch performance.
We highlight some comparisons on the real elapsed time required for which the training accuracy exceeds 0.8 in the case of $\epsilon =5$.
We obtained that the required times when setting $\lambda_\mathrm{target}$ to 0.1, 0.3, and 0.8 were approximately 270, 132, and 8 minutes, respectively. 
This comparison shows that the runtime performance with $\lambda_\mathrm{target} = 0.8$ is approximately 3.9 times faster than that with $\lambda_\mathrm{target} = 0.3$, and 8.0 times faster than that with $\lambda_\mathrm{target} = 0.1$.
Therefore, we would like to contend that the runtime performance can be improved significantly by setting $\lambda_\mathrm{target}$ to large (i.e., high transmission rate), when the path loss index $\epsilon$ is large.
\par
These results suggest that high-rate transmissions with sparse network topology will facilitate the development of efficient decentralized machine learning, especially in situations such as in indoor or urban channels. 

\section{Conclusion}
\label{sect:conclusion}
We proposed a novel communication strategy for D-PSGD on wireless systems by incorporating influences of the network topology.
We found that the influence of network density on the training accuracy of D-PSGD is {\it less} significant. Based on this finding, we designed the communication strategy for D-PSGD, in which each node communicates with a high-rate transmission rate under the constraint of the network density. This strategy enables to improve the runtime performance of D-PSGD while retaining high training accuracy. 
\par
Numerical evaluations showed that the network topology (transmission rate) highly influences on the runtime performance, especially in situations where the path loss index is large. 
We would like to conclude that the influences of the network topology will be a crucial factor that should be non-negligible to perform decentralized learning in wireless systems effectively, especially in indoor or urban scenarios.
\par
In future work, we will develop sophisticated optimization methods for the proposed strategy that can be applied to more complex situations such as those where location information is not available.

\section*{Acknowledgements}
This research was funded by The Telecommunications Advancement Foundation and the Japan Society for the Promotion of Science through KAKENHI under Grant 19K14988.

\bibliographystyle{IEEEtran}
\bibliography{reference}

\end{document}